\documentclass[a4paper]{jpconf-kb}
\usepackage{amsmath}
\usepackage{iopams}
\usepackage{natbib}
\usepackage{graphicx}

\usepackage{imakeidx}
\makeindex[name=subject,title=Subject index,columns=1]
\makeindex[name=authors,title=Author index,columns=1]

\def\be{\begin{equation}}
\def\ee{\end{equation}}
\newcommand{\ba}{\begin{eqnarray}}
\newcommand{\ea}{\end{eqnarray}}

\begin{document}

\title{Cosmic superstring networks with Y-junctions: Evolution, B-modes and Gravitational waves}

\author{D.A.\ Steer$^1$}

\address{$^1$ APC (UMR 7164 - APC, Univ Paris Diderot, CNRS/IN2P3, CEA/lrfu, Obs de Paris, 
Sorbonne Paris Cit\'e, France), 10 rue Alice Domon et L\'eonie Duquet, 75205 Paris Cedex 13, France.
}

\ead{steer@apc.univ-paris7.fr}

\index[authors]{Steer, D.A.}

\begin{abstract}

We review the properties and evolution of strings networks containing both a spectrum of string tensions
as well as Y-junctions\index[subject]{Y junctions}, namely a point at which three different strings meet. Such
a situation is expected in cosmic superstring networks\index[subject]{cosmic super strings}, where the tension
of the different strings is determined by the fundamental string tension $\mu_F$ as well as the string coupling $g_s$.
We discuss evolution of such a network, and the proliferation of kinks as they travel through junctions. These
effects can leave observational signals on B-modes\index[subject]{B-modes} and gravitational waves\index[subject]{Gravitational waves}, which can then in turn constrain
the underlying parameters of the theory.

\end{abstract}

\section{Introduction}

The last decade has seen intense effort devoted to trying to embed the inflationary paradigm into a more fundamental theory ---
and particularly superstring theory.  In some of the resulting models (and most successfully in warped `brane inflation'  [\cite{Kachru:2003sx}]),\index[authors]{Kachru, S.}\index[authors]{Kallosh, R.}\index[authors]{Linde, A.}\index[authors]{Maldacena, J.M.}\index[authors]{McAllister, L.P.} when the universe reheats at the end of inflation --- thus starting up the big-bang --- a network of `cosmic superstrings' is expected to form.   These cosmic superstrings are stable line-like objects of cosmic size whose properties and interactions depend on the two unknown parameters of the underlying superstring theory: the fundamental string coupling $g_s$, and the fundamental string tension $\mu_F$.  In the same way as their field theory counterparts (namely cosmic strings, which have been studied since the pioneering paper [\cite{Kibble:1976sj}]\index[authors]{Kibble, T.W.B.}), cosmic superstrings interact with cosmic fluids and leave observational signatures, and hence through experiment one hopes to have a direct window onto physics at the very highest of energy scales.   Since different observables depend in different ways on the parameters of the theory $g_s$ and $\mu_F$, a combination of measurements could then determine these parameters. This is illustrated in figure 1 (taken from [\cite{Pourtsidou:2010gu}]\index[authors]{Pourtsidou, A.}\index[authors]{Avgoustidis, A.}\index[authors]{Copeland, E.J.}\index[authors]{Pogosian, L.}\index[authors]{Steer, D.A.}), which combines forecasted Cosmic Microwave Background (CMB) constraints with  ones from the expected stochastic background of Gravitational waves (GW) emitted by a cosmic superstring network.\footnote{
Note that by construction in these models, $\mu_F$ is much smaller than the naive expectation of $m_{s}^{2}$, where $m_{s}$ is the mass scale of string theory. Consequently, $G \mu_F$, the dimensionless number measuring the tension of cosmic superstrings in units of the Newton constant, $G$, can be lowered below the current observational bounds. }

\begin{figure}[!ht]
\includegraphics[width=18pc]{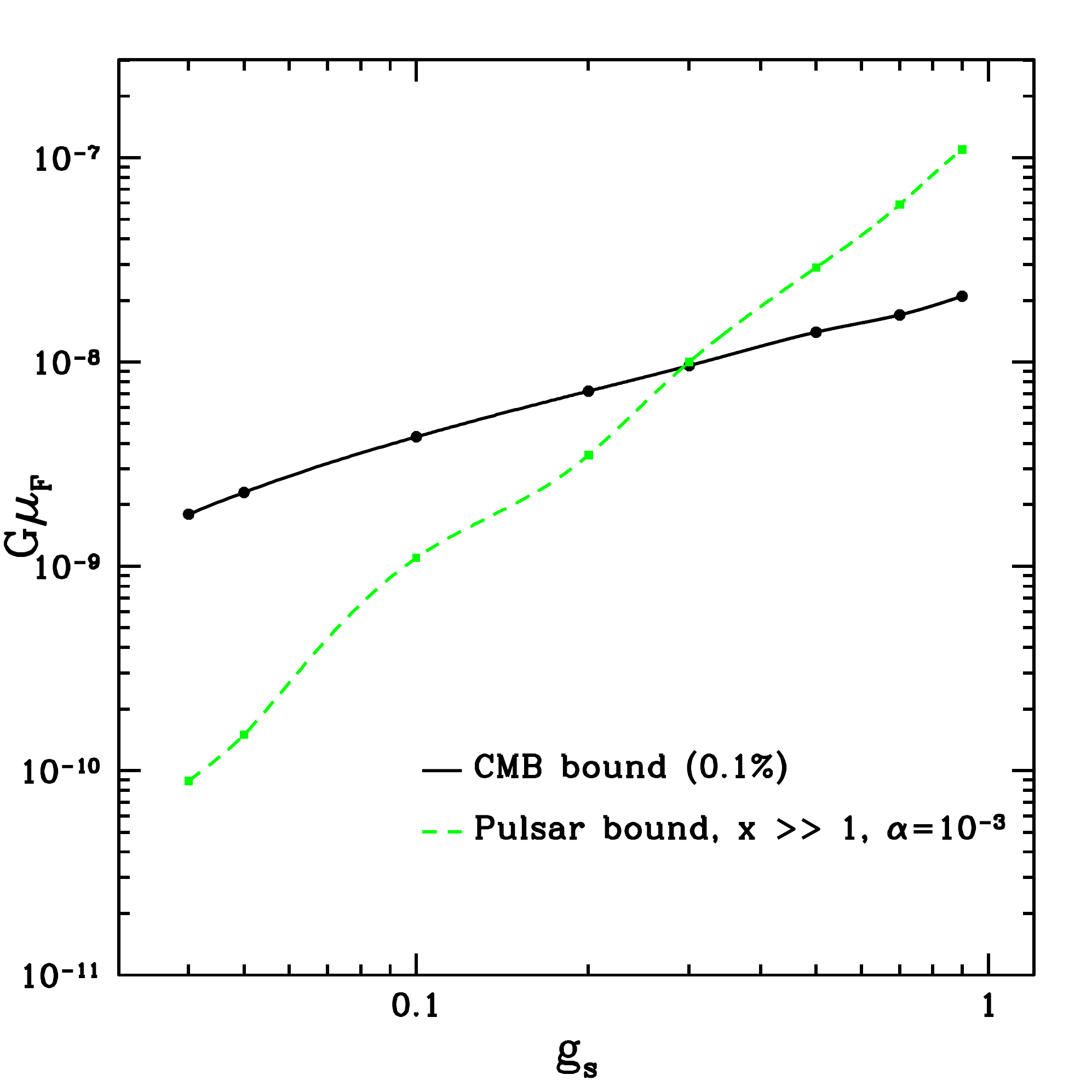}\hspace{2pc}%
\begin{minipage}[b]{17pc}\caption{\label{fig1}  The forecasted bound [\cite{Pourtsidou:2010gu}] on $\mu_F$ and $g_s$ from CMB measurements based on the hypothesis that cosmic superstrings
contribute $0.1\%$ to the CMB temperature anisotropies (solid-black line).  The bounds from pulsars  (short-dash-green line) on the stochastic background of GWs emitted from a network of cosmic superstrings (assuming `large loops').}
\end{minipage}
\end{figure}

In this paper we review some of the properties of cosmic superstrings, focusing in particular on the r\^ole of Y-junctions in the evolution of a tangle of such strings. By definition, a Y-junction\index[subject]{Y junctions} is a point at which two strings meet and bind together for form a third string.  These junctions are of course not unique to cosmic superstring networks but can also form in field-theory strings, and this analogy will be exploited in the following.
We also focus on two potentially important observational signatures of cosmic superstrings mentioned above --- B-modes and Gravitational waves.

\section{Local properties of cosmic superstrings}

Cosmic superstrings come in different varieties --- fundamental (F-) strings and D-strings (D1-branes), as well as $(p,q)$-strings (where $p$ and $q$ are integer coprime numbers) which are bound states of $p$ F-strings and $q$ D-strings (see e.g.~[\cite{Copeland:2003bj, Jackson:2004zg}]\index[authors]{Copeland, E.J.}\index[authors]{Myers, R.C.}\index[authors]{Polchinski, J.} \index[authors]{Jackson, M.G.}\index[authors]{Jones, N.T.}\index[authors]{Polchinski, J.}).   In the simplest case, namely when the extra dimensions present in superstring theory are assumed flat and compactified, the tension of a $(p_i,q_i)$ string is given by
\be\label{pqtension} 
 \mu_i \equiv  \mu_{(p_i,q_i)}=\frac{\mu_F}{g_s} \sqrt{p_i^2 g_s^2+q_i^2} \   \qquad (0\leq g_s \leq 1) \, .
\ee 
The lightest string is therefore the (1,0) string with tension $\mu_F$, and the next lightest is the (0,1) D-string with tension $\mu_F/g_s$.  Note that one always has
\be
\mu_i + \mu_j \geq \mu_{i+j} \, ,
\label{bound}
\ee
so that in particular the $(1,1)$ FD-string  is a bound state of the F- and D-strings.

\begin{figure}[h]
\centering
\includegraphics[scale=0.5]{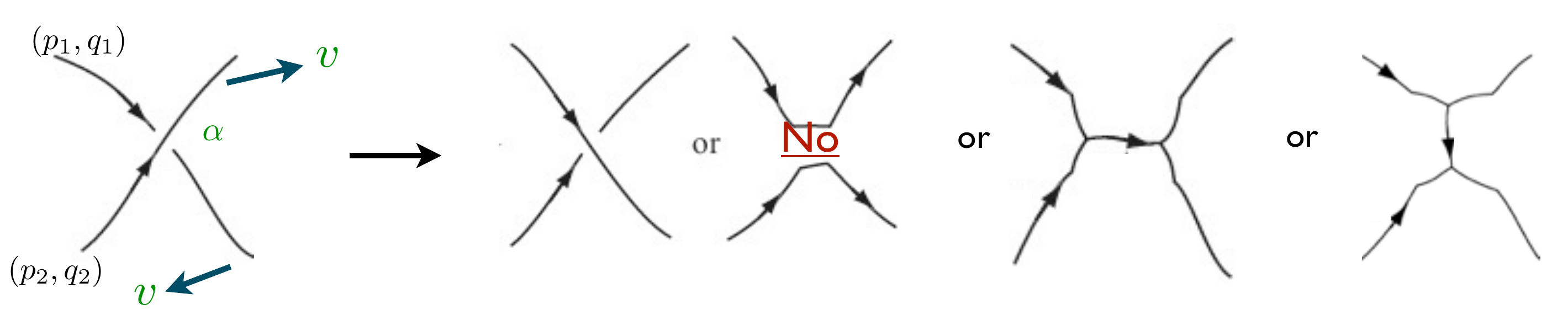}
\caption{\label{fig2} Possible outcomes of the collision between two cosmic strings of different types. Intercommutation is forbidden by charge conservation.}
\end{figure}

What is the outcome of a collision between two cosmic superstrings of different types?  To answer that question it is simplest to take an alternative view of a $(p,q)$-string with $q\neq 0$: indeed this can be seen as $p$ units of electric flux dissolved on the world volume of $q$ coincident D-strings. Since collisions must respect charge conservation, it thus follows that if a $(p_i,q_i)$-string and a $(p_j,q_j)$-string with $i\neq j$ collide, they can not intercommute.\index[subject]{string intercommutation} Instead they either go through each other, or exchange partners and form a Y-junction (see figure \ref{fig2}).  Notice from (\ref{bound}) that at the Y-junction the 3rd string is a bound state of the 2 colliding strings, and  since charges must be conserved at the junction $\sum_{i=1} q_i = 0 = \sum_{i=1}p_i$ [\cite{Copeland:2007nv}].\index[authors]{Copeland, E.J.}\index[authors]{Firouzjahi, H.}\index[authors]{Kibble, T.W.B.}\index[authors]{Steer, D.A.}

Y-junctions of course can also appear in networks of solitonic strings. The simplest example is that of Abelian-Higgs cosmic strings: in the Type-I regime cosmic strings of all windings $n \in Z$ are stable, and when two, say $n=1$ strings collide, they can form a junction (between the two $n=1$ strings and an $n=2$ string).  Junctions also exist in
Ônon-Abelian stringsÕ,\index[subject]{non abelian cosmic strings} for which the fundamental group $\pi_1({\cal M})$ of the manifold of degenerate vacua, which classifies the strings, is non-Abelian, see for e.g.~[\cite{Hindmarsh:1994re}].\index[authors]{Hindmarsh, M.}\index[authors]{Kibble, T.W.B.}  In both these cases, the strings are uncharged so there is no analogue of charge conservation, and such solitonic strings {\it may} intercommute.

The dynamics of 3 strings meeting at a junction can be determined from a generalised Nambu-Goto action\index[subject]{Nambu Goto action}.  Let $\mu_i$ $(i =1,2,3)$ denote the tensions of the strings, whose coordinates are $x^\mu_i (\sigma^a)$, and which evolve in a space-time with metric $g_{\mu \nu}$. (Here the two worldsheet coordinates are $\sigma^a=(\tau,\sigma)$, and $\mu=0,1,2,3$ is a space-time index).  Then the action for the system is given by [\cite{Copeland:2006eh}]\index[authors]{Copeland, E.J.}\index[authors]{Kibble, T.W.B.}\index[authors]{Steer, D.A.}
\ba
\label{action}
S= -\sum_{i} \mu_i   \int  {\rm d}  \tau  {\rm d} \sigma  \theta \left(s_{i}(\tau)-\sigma \right)  \sqrt{ -\det\left({g_{\mu \nu} x^\mu_{i,a} x^\nu_{i,b} }\right)}    
+ \sum_{i}  \int {\rm d}   \tau    f_{i \mu}  \left[  x^\mu_i(s_i (\tau),  \tau )-\bar x^\mu(\tau)  \right]
  \, ,
\ea
where the first term is Nambu-Goto action for each of the 3 strings, whilst the last term imposes --- through the Lagrange multiplier $f_i^\mu(\tau)$ --- that the 3 strings meet at the same point $\bar{x}^\mu(\tau)$. This is the position of the 3 strings when $\sigma=s_i(\tau)$.  In flat space-time $g_{\mu \nu}=\eta_{\mu \nu}$, and on imposing the conformal temporal gauge, the resulting equations of motion have been derived in [\cite{Copeland:2006eh}]\index[authors]{Copeland, E.J.}\index[authors]{Kibble, T.W.B.}\index[authors]{Steer, D.A.} and they can be solved given suitable initial conditions (namely the positions and velocities of the strings at the initial time).

\begin{figure}[!ht]
\begin{minipage}{18pc}
\includegraphics[width=18pc]{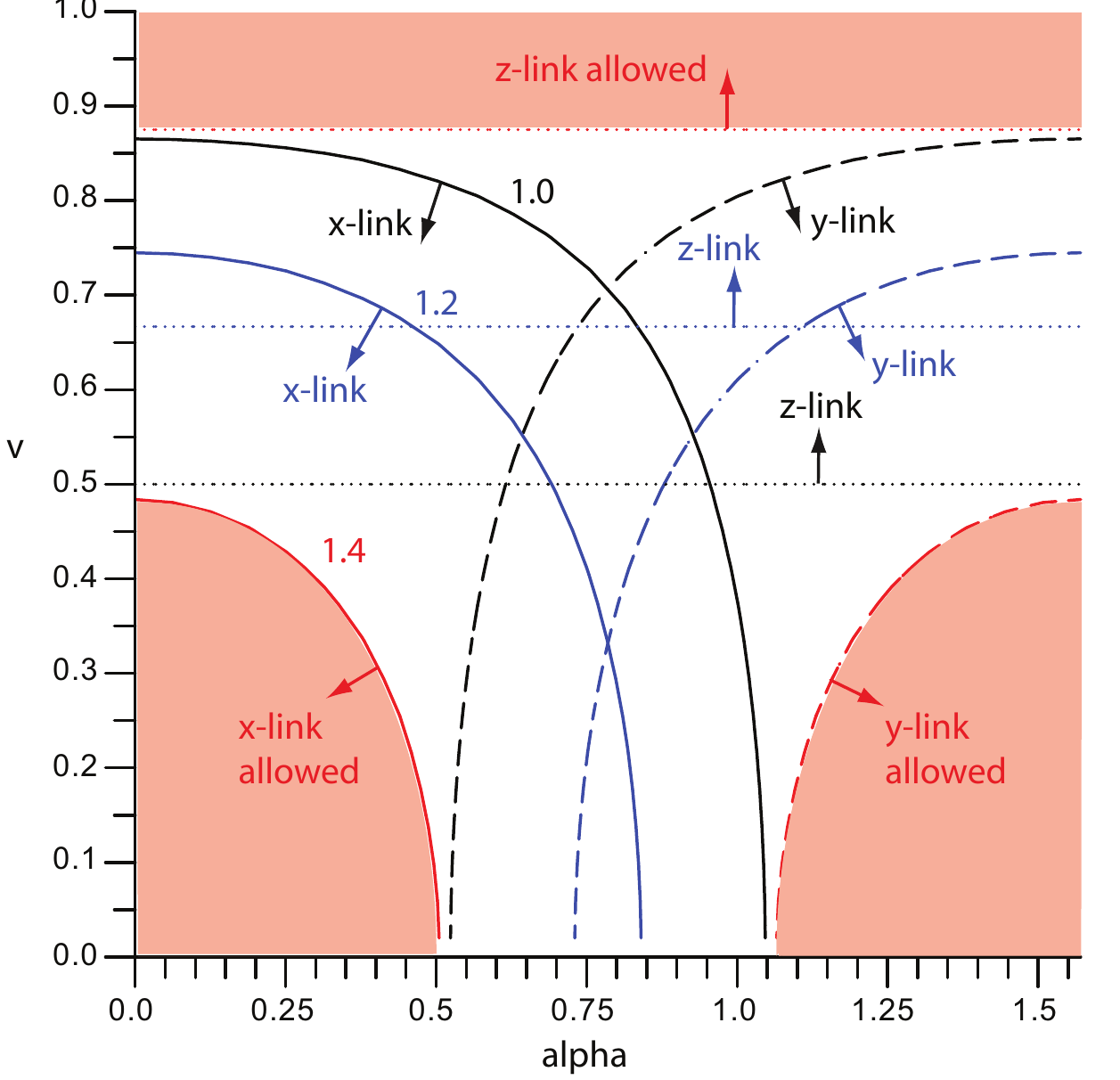}
\caption{\label{fig3} Kinematic constraints for $\mu_1=\mu_2$.  Allowed regions are to the left of the full curves for $x$-links; and for $y$-links to the right of the dashed curves. For $z$-links can form when non-abelian strings collide. The values of $\mu_3$ are 1.4 (red), 1.2 (blue), 1.0 (black).  Allowed regions are shaded for the $\mu_3=1.4$ case.}
\end{minipage}\hspace{2pc}%
\begin{minipage}{18pc}
\includegraphics[width=18pc]{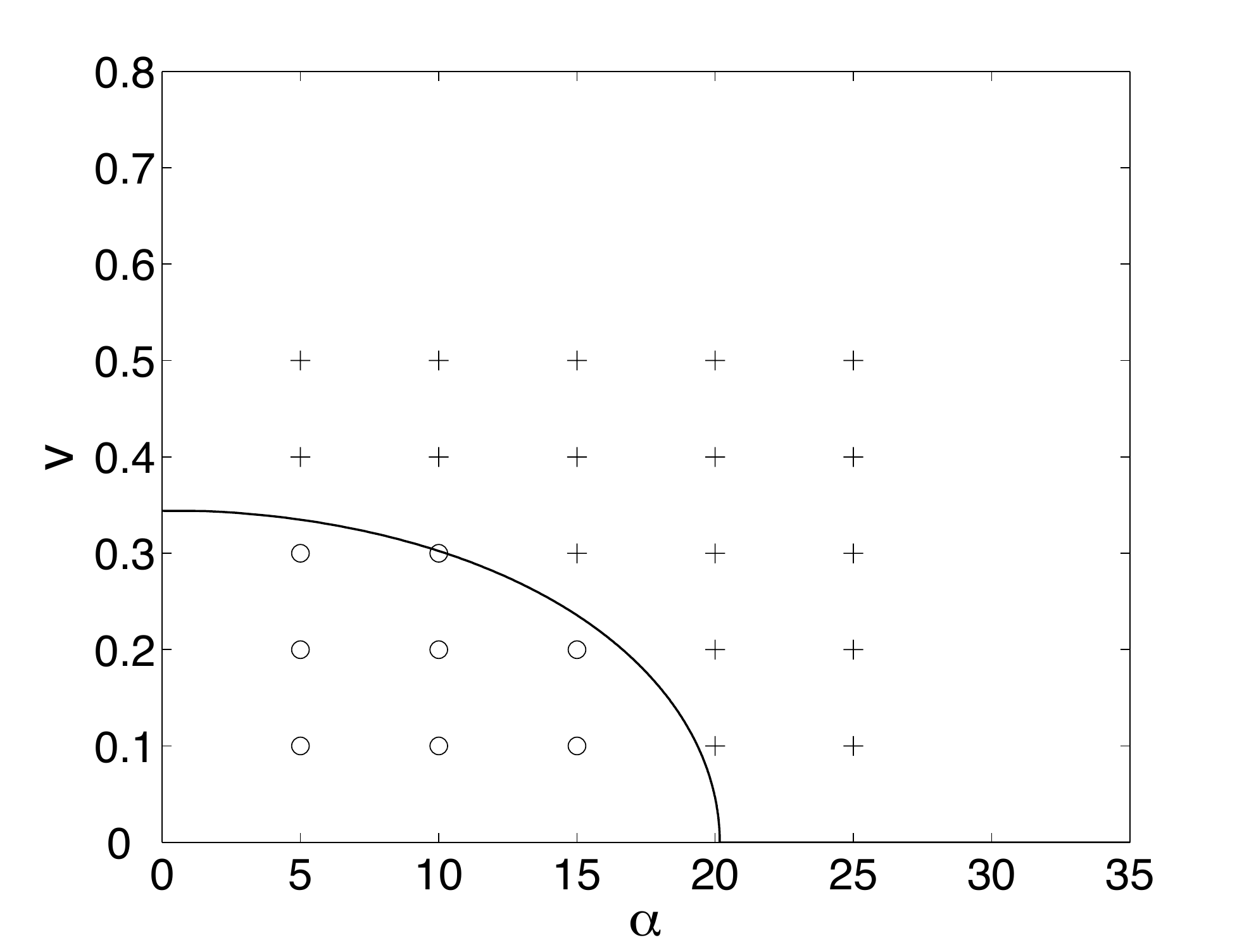}
\caption{\label{fig4} Results from a simulation [\cite{Salmi:2007ah}]\index[authors]{Salmi, P.}\index[authors]{Achucarro, A.}\index[authors]{Copeland, E.J.}\index[authors]{Kibble, T.W.B.}\index[authors]{de Putter, R.}\index[authors]{Steer, D.A.} of the collision of two winding 1 strings
in the type I Abelian-Higgs model (with ratio of square of the Higgs to vector masses given by $\beta=0.36$).
The solid line shows the theoretical curve of equality in~(\ref{inequality}). The
+ mark the events of single intercommutation, whereas $\circ$ indicates 
strings merging together to form an $x$-link, a string with 
winding number $n=2$ (as observed in the simulation).}
\end{minipage} 
\end{figure}

In particular, these equations of motion can be used to determine the region of phase space in which the different collision outcomes in figure \ref{fig2} can occur.  Indeed, as in that figure, consider a straight string of tension $\mu_1$ making an angle $\alpha$ with the positive $x$-axis, and another of tension $\mu_2$ at angle $-\alpha$. Both strings are in the $xy$-plane and have a velocity $v$ along the $z$-axis with opposite directions. Once they intersect, they can potentially form a third string, an Òx-linkÓ which has tension $\mu_3$ (we follow here the notation introduced in [\cite{Copeland:2006if}];\index[authors]{Copeland, E.J.}\index[authors]{Kibble, T.W.B.}\index[authors]{Steer, D.A.} if $\mu_1=\mu_2$ then the x-link is indeed positioned along the $x$-axis) or a ``y-link''.  On using the equations of motion derived from (\ref{action}) for this configuration,
kinematic constraints on the parameters $(\alpha,v)$ follow from the requirement that the total length of the progeny string (linking the 2 junctions) must {\it increase}.  When $\mu_2=\mu_1$, this requirement  takes the simple form
\be
\alpha < {\rm arccos}\left(\frac{\gamma \mu_3}{2\mu_1} \right)
\label{inequality}
\ee
where $\gamma = 1/\sqrt{1-v^2}$, and is shown in figure \ref{fig3}.  Notice that no ``x-link'' is allowed if the velocity of the strings is too large.  Numerical experiments with type I abelian-Higgs vortices have shown that when junction formation is kinematically allowed, it dynamically takes place, see figure \ref{fig4}.

\section{Network evolution}

The equations of motion derived from (\ref{action}), together with the kinematic constraints, can be used to study network dynamics.

\begin{figure}[!ht]
\begin{minipage}{16pc}
\includegraphics[width=16pc]{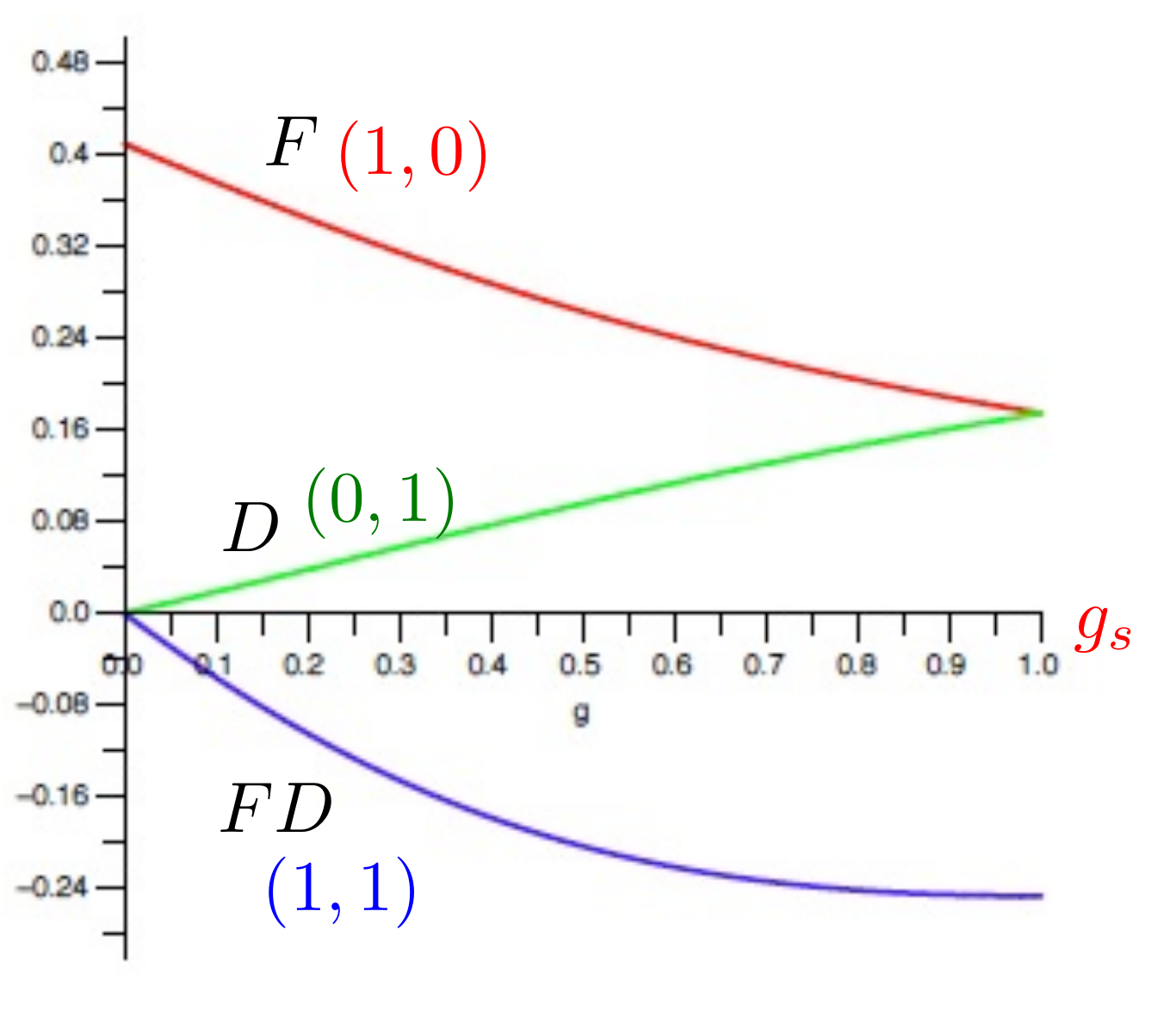}
\caption{\label{fig5} Average rate 
of increase of length of F-, D- and FD-strings, as a function of $g_s$, predicted
from action (\ref{action}) in {\it flat-space}.
}
\end{minipage}\hspace{2pc}%
\begin{minipage}{16pc}
\includegraphics[width=16pc]{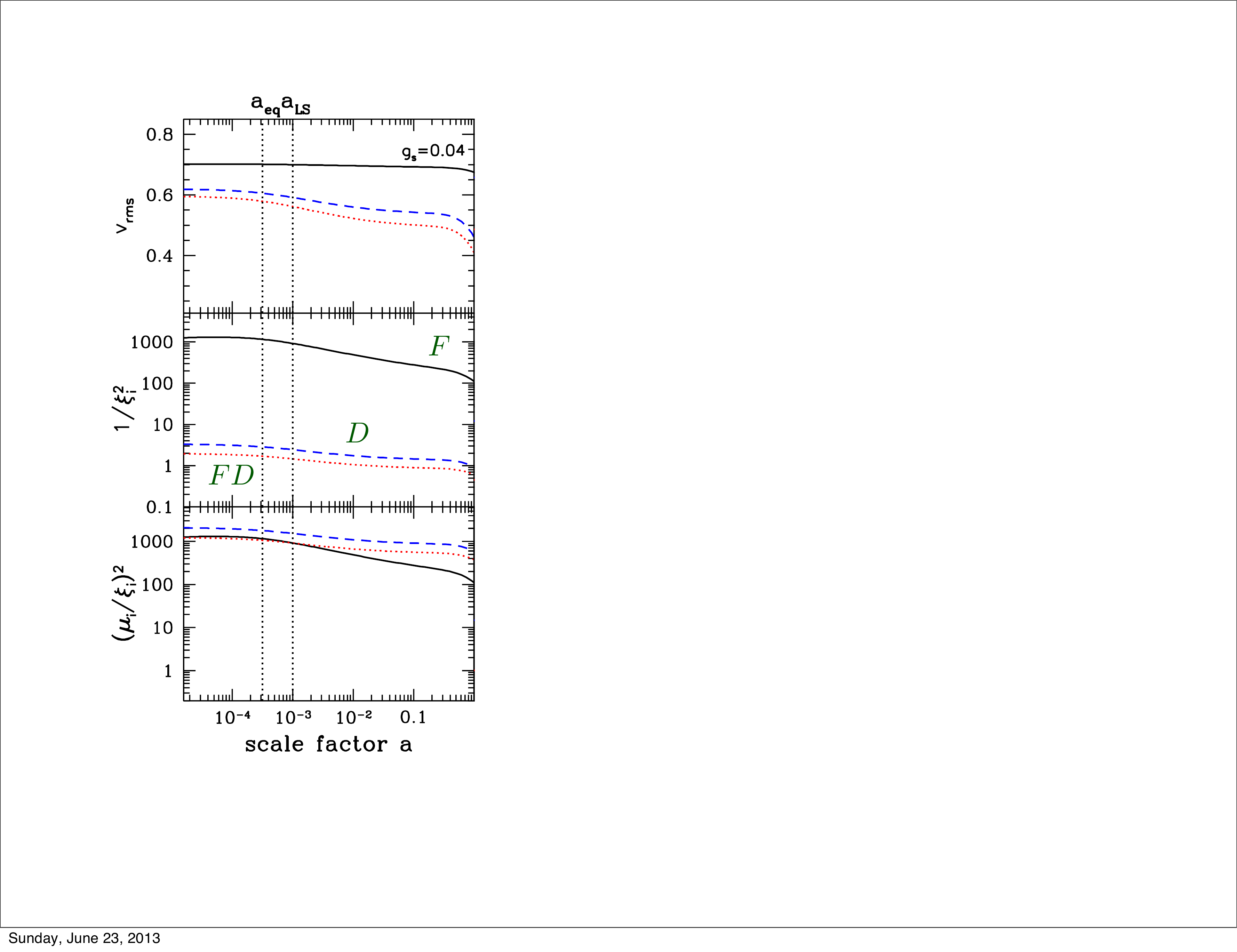}
\caption{\label{fig6} Taking into account energy loss, Hubble expansion, and network dynamics: evolution of the rms velocity $v_i$, number density $\xi_i^{-2} \equiv (t/L_i)^2$, and the power spectrum density $(\mu_i/\xi_i)^2$.}
\end{minipage} 
\end{figure}

As a first step, let us ignore Hubble expansion and any energy loss mechanisms. Then the energy in the string network is fixed, but some strings will shorten and others will grow: how fast on average, is this growth or shortening; and how fast on average do the junctions move along a string of tension $\mu_i$? These questions, and others, were addressed in [\cite{Copeland:2007nv,Copeland:2006if}],\index[authors]{Copeland, E.J.}\index[authors]{Kibble, T.W.B.}\index[authors]{Steer, D.A.} on assuming that at any Y-junction, the strings are straight with their incoming unit vectors randomly distributed on the unit sphere.  Figure \ref{fig5} shows the average rate of increase of cosmic superstrings of type $i$, as a function of $g_s$, where we have solely focused on F-, D- and FD-strings.   From (\ref{pqtension}) for $g_s \ll 1$ it follows $\mu_F \ll \mu_D \sim \mu_{FD}$. Furthermore from figure \ref{fig5}, we see that the rate of increase in length of the F-strings is the largest. We can therefore expect these to be the most populous strings in the network. Furthermore, one can ask if they are sufficiently populous to also dominate the energy density, despite their small tension --- we will return to this point below (see also [\cite{Rivers:2008je}]\index[authors]{Rivers, R.}\index[authors]{Steer, D.A.}  who have studied the statistical mechanics of string networks with junctions).   When $g_s \simeq 1$, then $\mu_F \sim \mu_D \sim \mu_{FD}/\sqrt{2}$ and one expects similar numbers of $F$ and $D$ strings.

To understand cosmological network dynamics fully, however, it is crucial to include Hubble expansion and loop formation. Furthermore, cosmic superstrings are intrinsically quantum objects which are not strictly localised in 4D --- they can fluctuate and explore the extra dimensions, where there is sufficient `space' for them to miss each other when they go past.  To take this into account, the classical equations of motion coming from action (\ref{action}) (as well as the kinematic constraints discussed above) must be augmented to account for the probability that string $i$ interacts with string $j$ in the first place. This so-called `intercommuting probability' $P_{ij}$ is $(v,\alpha,g_s$)-dependent, and also depends on a further parameter $w$, which is related to the volume of the extra dimensions. The explicit expressions can be found in for e.g.~[\cite{Pourtsidou:2010gu}]\index[authors]{Pourtsidou, A.}\index[authors]{Avgoustidis, A.}\index[authors]{Copeland, E.J.}\index[authors]{Pogosian, L.}\index[authors]{Steer, D.A.} for different string types.  Note that for field theory strings this probability is (essentially always) equal to 1, whereas for cosmic superstrings it can be much smaller than 1.

Given the form of $P_{ij}$, the kinematic constraints, and the classical equations of motion, one can then attempt to model the network evolution --- in [\cite{Pourtsidou:2010gu}]\index[authors]{Avgoustidis, A.}\index[authors]{Copeland, E.J.}\index[authors]{Pogosian, L.}\index[authors]{Steer, D.A.}  we have use a generalised `velocity dependent 1-scale model' (this was first developed to describe a tangle of type II Abelian-Higgs cosmic strings with no junctions, and only strings of a single tension [\cite{Martins:1996jp}]).\index[authors]{Martins, C.J.A.P.}\index[authors]{Shellard, E.P.S.S.}   The variables describing the bulk properties of the system are taken to be {(i)} the energy density $\rho_i$ in infinite strings of type $i$, from which one defines a correlation length $\rho_i \equiv \mu_i/L_i^2$, and {(ii)} the rms velocity $v_i$ of the string of type $i$.  These satisfy the following coupled set of equations (for simplicity we assume that all the energy liberated by the formation of junctions is radiated away, and the dot denotes ${\rm d}/{\rm d} t$):
\be\label{rho_idtgen} 
    \dot\rho_i = -2\frac{\dot a}{a}(1+v_i^2)\rho_i-\frac{c_i 
    v_i\rho_i}{L_i} - \sum_{a,k} \frac{d_{ia}^k   
    \bar v_{ia} \mu_i \ell_{ia}^k(t)}{L_a^2 L_i^2} + \sum_{b,\,a\le b}   
    \frac{d_{ab}^i \bar v_{ab} \mu_i   
    \ell_{ab}^i(t)}{L_a^2 L_b^2}         \, ,
  \ee  
  \be
 \dot v_i = (1-v_i^2)\left[\frac{1}{L_i}\left(\frac{2\sqrt{2}(1-8v_i^6)}{\pi(1+8v_i^6)}\right)-2\frac{\dot a}{a}v_i \right] \, \, , \label{v_idtgen}     
\ee
where the presence of Y-junctions is reflected in the last two terms in (\ref{rho_idtgen}); for instance, the penultimate one models the energy loss from the strings of type $i$ due to their collision with strings of type $a$ (with average relative velocity $\bar{v}_{ia}$), resulting in the formation of type $k$ links of average length $\ell_{ia}^k(t)$.  This occurs with $g_s$-dependent probability $d_{ia}^k=d_{ai}^k$ which depends on the ${\cal P}_{ia}$ and kinematic constraints. Self-interactions of strings of type $i$, leading to the formation of loops and thus the removal of energy from the long string network of type $i$, are quantified by the coefficients $c_i$ in (\ref{rho_idtgen}) which now depend on ${\cal P}_{ii}$ and are thus also $g_s$-dependent.
(Again see  [\cite{Pourtsidou:2010gu}]\index[authors]{Avgoustidis, A.}\index[authors]{Copeland, E.J.}\index[authors]{Pogosian, L.}\index[authors]{Steer, D.A.} for explicit expressions for the different coefficients.)

%
%

\begin{figure}[!ht]
\begin{minipage}{18pc}
\includegraphics[width=18pc]{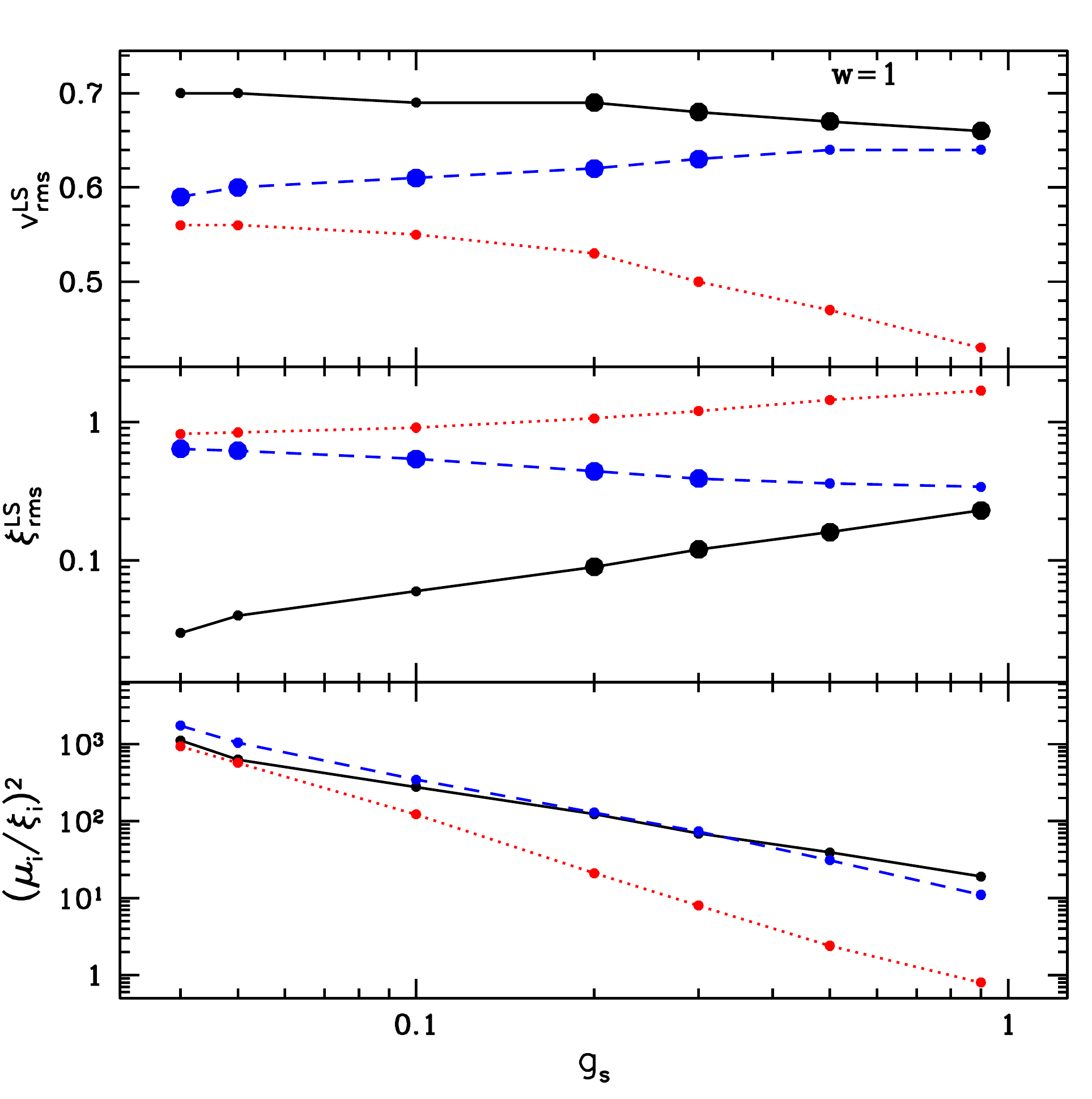}
\caption{\label{fig7}  Dependence of the $v_{i}$, $\xi_i$, and $M_i$ at the time of last scattering (LS) as a function of $g_s$.  Labeling corresponds that of figure 6. The string type(s) dominating the power spectrum are shown with oversized dots.
}
\end{minipage}\hspace{2pc}%
\begin{minipage}{18pc}
\includegraphics[width=18pc]{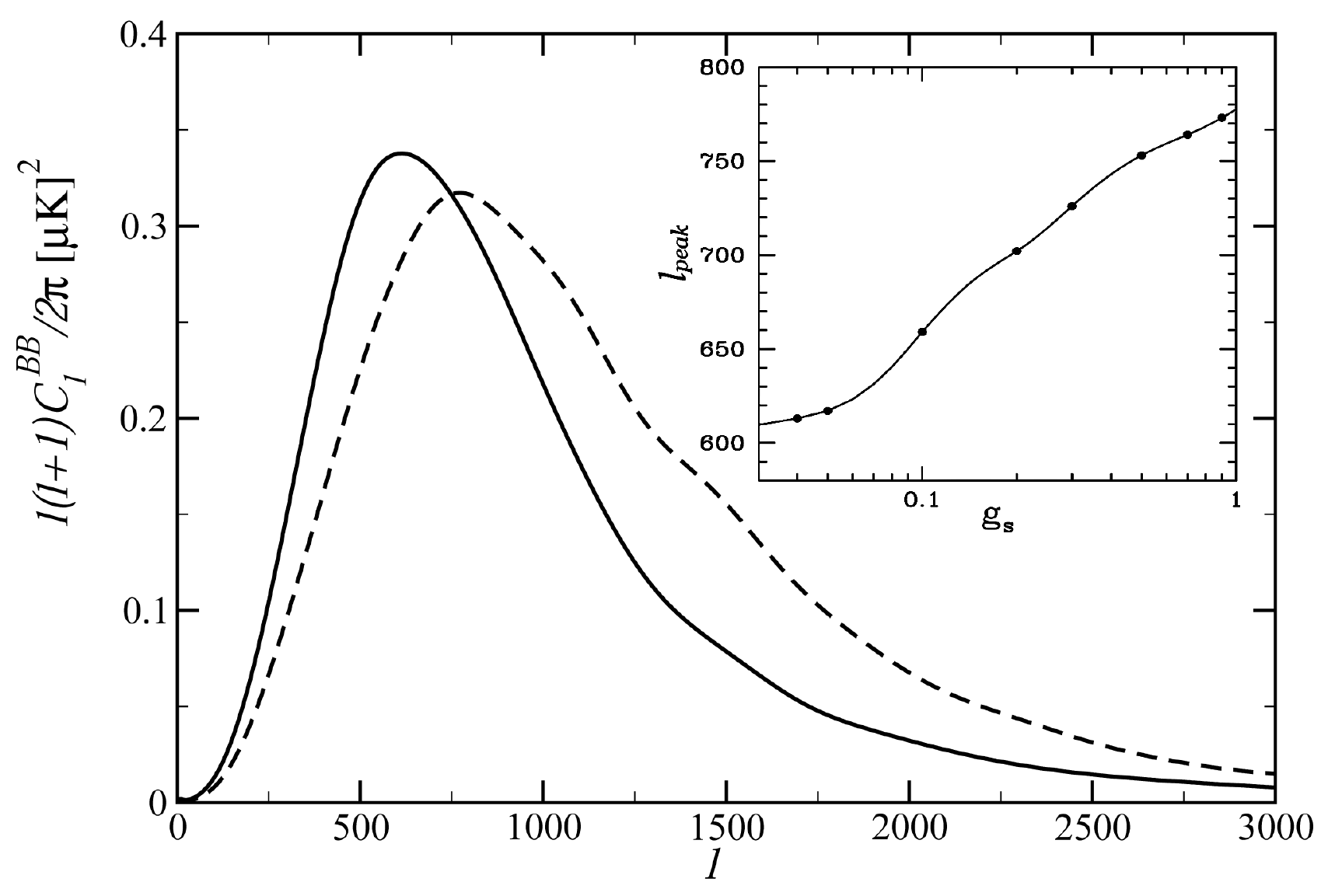}
\caption{\label{fig8} The B-mode power spectra for $g_s=0.04$ (solid lines) and $g_s=0.9$ (dashed lines) normalised to $f_s=0.1$. The insert plot shows the position of the peak as a function of $g_s$.}
\end{minipage} 
\end{figure}

We have solved Eq.~(\ref{rho_idtgen}-\ref{v_idtgen}) numerically for different values of our external parameters $g_s$ and $w$ (and for the 7 lightest strings in the hierarchy given in (\ref{pqtension})).  
The system is found to generically reach an attractor \emph{scaling}  solution in which  $\xi_i\equiv L_i/t$ and $v_i$ asymptotically approach constant values during the radiation and matter eras.  For all studied values of the string coupling $g_s$, the three lightest strings (F, D, and FD), dominate the two relevant physical quantities --- {\it string number density}
$N_i = \xi_i^{-2}$, and the {\it power spectrum density} $M_i\equiv \left({\mu_i}/{\xi_i}\right)^2$ of the network.
These are shown in figure \ref{fig6} together with $v_i$ for $g_s=0.04$. Just as expected from the discussion above, the lightest F-string is by far the most populous.  However, regarding 
the energy density in the 3 different types of strings, the large number of F-strings does not outweigh their light tension: in fact, the heavier and less numerous D strings are seen to dominate the power spectrum density. 
Finally figure \ref{fig7} shows the dependence of $v_i$, $\xi_i$ and $M_i$ on $g_s$ {\it at the time of last scattering}. These quantities are crucial to determining the shape of the B-mode power-spectrum.
%


\section{B-modes}

We have examined the imprint of cosmic superstrings on the CMB temperature and polarisation spectra, as a function of $g_s$ and $\mu_F$, by modifying the publicly available code CMBACT
so as to allow for strings with multiple tensions whose scaling is modelled by Eqs.~(\ref{rho_idtgen}) and (\ref{v_idtgen}), see [\cite{Pourtsidou:2010gu,Avgoustidis:2011ax}].\index[authors]{Avgoustidis, A.}\index[authors]{Copeland, E.J.}\index[authors]{Pogosian, L.}\index[authors]{Steer, D.A.}\index[authors]{Moss, A.}

The crucial point to note is that even with a marginal contribution to the TT spectrum of $1\%$, strings can be a {\it prominent source} of B-mode polarisation  on subdegree scales since strings, unlike inflation, are actively sourcing {\it vector} mode perturbations of magnitude comparable to the scalar perturbations (see for instance [\cite{Pogosian:2007gi}]\index[authors]{Pogosian, L.}\index[authors]{Wyman, M.} and references within).  
The shapes of the spectra depend on the large-scale properties of the network, namely $\xi_i$ and $v_i$. In particular, the position of the main B-mode peak moves towards lower $\ell$ for larger correlation length, while the dependence on the rms velocity is nonlinear.

The amplitude of the CMB angular spectra $C_{\ell}$ for multi-tension networks 
is proportional to the sum of the power spectrum densities $M_i$ of each string type: $C_\ell^{strings} \propto M_{\rm total} = \sum_{i=1}^{N} M_i$. This, together with observational data, ultimately constrains the fundamental string tension $\mu_F$, and we have adjusted it such that
$f_s=C^{TT}_{strings} / C^{TT}_{total} \leq 0.1$, where $C^{TT}_{strings}$ and $C^{TT}_{total}$ are defined by summing 
the corresponding $C_\ell$'s over multipoles $2\le\ell\le2000$.
 Figure \ref{fig1} shows the constraint on $\mu_F$ as a function of $g_s$ assuming $f_s=0.1$ (post-Planck, this number should be reduced).

 Figure \ref{fig8}, on the other hand, 
shows the B-mode spectra for the limiting values $g_s=0.04$ and $g_s=0.9$.  We see that the peak shifts significantly towards low multipole number as $g_s$ decreases.  The reason for this is the dependence of the BB peak position on $\xi_i$ and $v_i$  for the string type $i$ dominating the power spectrum density $M_i$.  Figure \ref{fig7} shows that the velocity of the dominant string type only varies between $0.6$ and $0.7$, so the peak position should be determined mostly by the dominant correlation length $\xi_i$.  As we have seen in figure \ref{fig6}, $M_i$ is dominated by F strings for large $g_s$, but at lower $g_s$ there is a transition and the rare (larger $\xi_i$), heavy D strings become more important.  As $g_s$ decreases, the D-string correlation length increases (Fig.~\ref{fig7}, middle panel) and the peak moves further down in $\ell$ (see Fig.~\ref{fig8}). 

Our result implies that an observation of the B-mode peak could help to break the degeneracy between $\mu_F$ and $g_s$ in the constraint imposed by the normalisation condition $f_s < 0.1$.  This degeneracy may also be broken, figure \ref{fig1}, by combining CMB with pulsar timing constraints (see below).  Finally, the reader is referred to [\cite{Avgoustidis:2011ax}]\index[authors]{Avgoustidis, A.}\index[authors]{Copeland, E.J.}\index[authors]{Pogosian, L.}\index[authors]{Steer, D.A.} for a discussion of whether the next generation of B-mode experiments will have the sensitivity to detect the peak, and hence the means to pin down the value of the string coupling constant, a direct observational test of string theory: we feel optimistically positive!

\section{Kink proliferation and Gravitational Waves}

It is well known that networks of cosmic strings generate a stochastic background of Gravitational Waves (GWs).  The reason for this is that a closed loop of length $\ell$ (with no junctions) oscillates periodically with fundamental period $T=\ell/2 \equiv 2\pi/\omega$, and emits GWs with frequency $\omega_n = n \omega$ ($n \in Z$).  As a result the loop decays in a lifetime $\tau= \ell/(\Gamma G\mu)$ where numerical simulations have shown that $\Gamma \sim 50$.  Emission is generally dominated by the lowest mode numbers $n$, and summing over all GWs emitted by  the numerous loops formed by the network during the course of its evolution (and taking into account the redshift effects) gives rise to a stationary and nearly Gaussian stochastic GW background. (The reader is referred to [\cite{Binetruy:2012ze}]\index[authors]{Binetruy, P.}\index[authors]{Bohe, A.}\index[authors]{Caprini, C.}\index[authors]{Dufaux, J.-F.} for a detailed review).  The stochastic GW background is characterised by
\be
h^2 \Omega_{GW}(f) = \left(\frac{h^2}{\rho_c}\frac{{\rm d}\rho_{GW}}{{\rm d} \ln f}\right)_{\rm today}
\ee
where $h$ is the uncertainty in the Hubble parameter, and $\rho_c$ the critical density today. It is constrained by pulsar timing experiments:  $h^2 \Omega_{GW} < 2 \times 10^8$ at $f_{\rm pulsar} = 1/8 {\rm yr}^{-1}$ [\cite{Jenet:2006sv}].\index[authors]{Jennet, A.}\index[authors]{Hobbs, G.}\index[authors]{van Straten, W.}\index[authors]{Manchester, W.}\index[authors]{Bailes, M.}  Figure 1 shows this pulsar bound for cosmic superstrings, where we have only considered loops of F-, D- and FD-strings, and all loops are assumed to contain {\it no} Y-junctions.  The $\mu_F$ and $g_s$-dependence of the constraint shown in figure 1 arises from the dependence of the string tensions on these parameters (Eq.~(\ref{pqtension})), as well as that of $c_j$ (the loop chopping efficiency in Eq.~(\ref{rho_idtgen})): as explained above, this latter parameter determines how many loops of type $j$ are formed at time $t$.\footnote{Note that all predictions of GW from cosmic strings suffer from important uncertainties, namely the typical size
 of a loop when it is chopped off the network, and the loop distribution.  These uncertainties does not affect the CMB predictions which are determined by the large scale evolution and on which there is much better agreement.}

In [\cite{Damour:2000wa}]\index[authors]{Damour, T.}\index[authors]{Vilenkin, A.} it was pointed out that previous estimates of the stochastic GW background may have been too large, since kinks (discontinuities on the tangent vector of a string) and cusps (points where the string instantaneously goes at the speed of light) lead to strong GW bursts --- namely short bursts of radiation which, in the case of kinks, are emitted in a fan-shaped set of directions.   These infrequent bursts should not be included in the calculation of the stochastic GW background.  However, an exciting idea is that their characteristic features may provide a new signature for detecting cosmic strings in the sky.

\begin{figure}[h]
\centering
\includegraphics[scale=0.5]{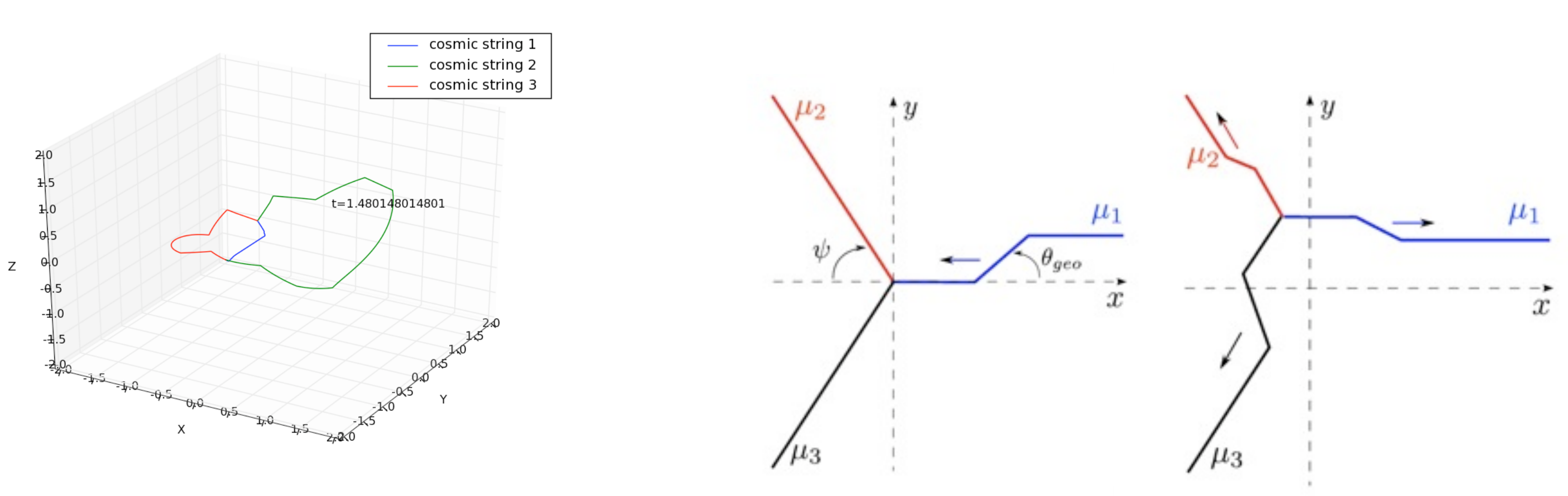}
\caption{\label{fig9} LH panel: a loop with junctions.  RH panel: detail of the propagation of a kink through a junction. Here two kinks initially propagate towards the junction, where they are reflected and transmitted to produce 6 final kinks.}
\end{figure}

Regarding this point, we finish by highlighting a potentially new source of GW bursts unique to string networks with junctions. Though not included explicitly in the equations (\ref{rho_idtgen})-(\ref{v_idtgen}), these networks can also chop off closed loops of strings containing (at least 2) Y-junctions, an example of which is shown in figure \ref{fig9}.  One can study the dynamics of such loops using the equations of motion coming from (\ref{action}), and the important observation is that when a kink arrives at a junction, not only is it reflected, but two transmitted kinks are also formed [\cite{Copeland:2006eh}].\index[authors]{Copeland, E.J.}\index[authors]{Kibble, T.W.B.}\index[authors]{Steer, D.A.} From 1 initial kink, we therefore go to 3 kinks, and hence the number of kinks on a closed evolving loops with junctions increases exponentially with time [\cite{Binetruy:2010bq}].\index[authors]{Binetruy, P.}\index[authors]{Bohe, A.}\index[authors]{Hertog, T.}\index[authors]{Steer, D.A.}  The amplitude of the GW burst emitted by a kink is proportional to the sharpness of the kink, and one can show that the number of kinks with amplitude larger than any fixed value also increases exponentially with time.  Indeed, following simulations of realistic loops, it is not unusual to find $k \gtrsim 10^4$ kinks of sharpness ${\cal O}(1)$ [\cite{Binetruy:2010bq}].\index[authors]{Binetruy, P.}\index[authors]{Bohe, A.}\index[authors]{Hertog, T.}\index[authors]{Steer, D.A.}

What is the effect of these kinks on the GW signal emitted by a network? Let $q$ denote the fraction of loops with junctions relative to loops without junctions. Consistent with a scaling solution, we assume $q$ is constant, and we take $qk \gg 1$.  (Furthermore, for simplicity, we consider a network in which all strings have the {\it same} tension, but are allowed to meet a junctions.)  One might expect that this large number of kinks might make it easier to detect bursts from string networks with junctions --- however, this is not always the case [\cite{Binetruy:2010cc}].\index[authors]{Binetruy, P.}\index[authors]{Bohe, A.}\index[authors]{Hertog, T.}\index[authors]{Steer, D.A.}  Clearly, in order for individual bursts to be resolved by a detector (eg LISA), they must arrive separately in that detector. The large number of kinks on a loop necessarily means that the burst event rate is very large, and if it is too high, the bursts start to superimpose instead of arriving one by one. Their random superposition contributes to the stochastic background of gravitational waves, and when $qk^2 \gg 1$ this is what happens [\cite{Binetruy:2010cc}]\index[authors]{Binetruy, P.}\index[authors]{Bohe, A.}\index[authors]{Hertog, T.}\index[authors]{Steer, D.A.} --- kink proliferation on loops with junctions generally makes individual bursts unobservable.\footnote{Note that we have assumed that $\Gamma$, the parameter determining the lifetime of a loop is of order 50. However, when there are many radiating kinks on a loop, one might expect $\Gamma$ to increase: see [\cite{Bohe:2011rk}]\index[authors]{Bohe, A.} for a detailed discussion of this point and its effect on GW emission.}

\section{Conclusions}

In this contribution we have highlighted some of the facets of strings with junctions, starting from their local properties (dynamics and the result of collisions), then discussing the evolution of a network of such strings, and finally considering their observational signatures on the CMB and in Gravitational Waves.  If cosmic superstrings really exist, and with a tension that is not too small, then through a combination of observations we conclude that it may be possible to open a window onto physics at the very highest of energy scales.

\ack

I am grateful to the organisers for the invitation to speak at this very interesting workshop. I would also like to thank all my different collaborators: it has been a great pleasure to work with them on the subjects presented here.

\bibliographystyle{jfm2}

\bibliography{Steer-bibliography}
\end{document}